\documentclass[runningheads]{llncs}

\usepackage{eccv}

\usepackage{eccvabbrv}

\usepackage{graphicx}
\usepackage{booktabs}

\usepackage[accsupp]{axessibility}  
\usepackage{algorithm}
\usepackage{algorithmicx}
\usepackage{algpseudocode}
\usepackage{wrapfig}
\usepackage{nicefrac}

\usepackage{hyperref}

\usepackage{orcidlink}
\usepackage{amsmath,amsfonts,bm}

\newcommand{\rvb}[0]{\mathbf{b}}

\newcommand{\rvu}[0]{\mathbf{u}}
\newcommand{\rvv}[0]{\mathbf{v}}

\newcommand{\rvx}[0]{\mathbf{x}}
\newcommand{\rvy}[0]{\mathbf{y}}

\newcommand{\mH}[0]{\bm{H}}
\newcommand{\mI}[0]{\bm{I}}

\newcommand{\mU}[0]{\bm{U}}
\newcommand{\mV}[0]{\bm{V}}

\newcommand{\mSigma}[0]{\bm{\Sigma}}

\newcommand{\E}{\mathbb{E}}

\newcommand{\normal}[2]{\mathcal{N}\left(#1,#2\right)}
\newcommand{\R}{\mathbb{R}}

\newcommand{\Cov}{\mathrm{Cov}}

\usepackage{amsmath}
\DeclareMathOperator*{\argmax}{argmax}
\DeclareMathOperator*{\argmin}{argmin}

\DeclareMathOperator*{\trace}{Tr}

\begin{document}

\title{Adaptive Compressed Sensing with Diffusion-Based Posterior Sampling} 


\author{Noam Elata\inst{1}\orcidlink{0009-0000-2692-2781} \and
Tomer Michaeli\inst{1}\orcidlink{0000-0003-0525-8054} \and
Michael Elad\inst{2}\orcidlink{0000-0001-8131-6928}}

\authorrunning{N.~Elata et al.}

\institute{Dept. of Electric and Computer Engineering \ \  \mbox{\and Dept. of Computer Science} \\
Technion - Israel Institute of Technology, Haifa, Israel \\
\email{\{noamelata@campus, tomer.m@ee, elad@cs\}.technion.ac.il}\\}

\maketitle

\begin{abstract}
    Compressed Sensing (CS) facilitates rapid image acquisition by selecting a small subset of measurements sufficient for high-fidelity reconstruction.  
    Adaptive CS seeks to further enhance this process by dynamically choosing future measurements based on information gleaned from data that is already acquired. 
    However, many existing frameworks are often tailored to specific tasks and require intricate training procedures.
    We propose AdaSense, a novel Adaptive CS approach that leverages zero-shot posterior sampling with pre-trained diffusion models.
    By sequentially sampling from the posterior distribution, we can quantify the uncertainty of each possible future linear measurement throughout the acquisition process.
    AdaSense eliminates the need for additional training and boasts seamless adaptation to diverse domains with minimal tuning requirements. 
    Our experiments demonstrate the effectiveness of AdaSense in reconstructing facial images from a small number of measurements.
    Furthermore, we apply AdaSense for active acquisition of medical images in the domains of magnetic resonance imaging (MRI) and computed tomography (CT), highlighting its potential for tangible real-world acceleration.\footnote{Our code is available at \url{https://github.com/noamelata/AdaSense}.}
\end{abstract}

\section{Introduction}

Compressed sensing (CS)~\cite{donoho2006compressed, candes2006compressive} methods have shown promise in image acquisition and reconstruction by leveraging the statistical prior of real-world signals. This allows capturing significant information using fewer measurements, which is crucial when acquisition is time-consuming or costly. To further optimize this process, adaptive CS techniques~\cite{ji2008bayesian, seeger2011large, arias2012fundamental, malloy2014near, nakos2018improved} use reconstructions based on existing measurements to dynamically prioritize subsequent ones. Both adaptive and non-adaptive CS have benefited from the recent surge in deep learning research~\cite{jalal2020robust, van2018compressed, wu2019deep}.

\begin{figure}[tb]
  \centering
  \includegraphics[width=\textwidth]{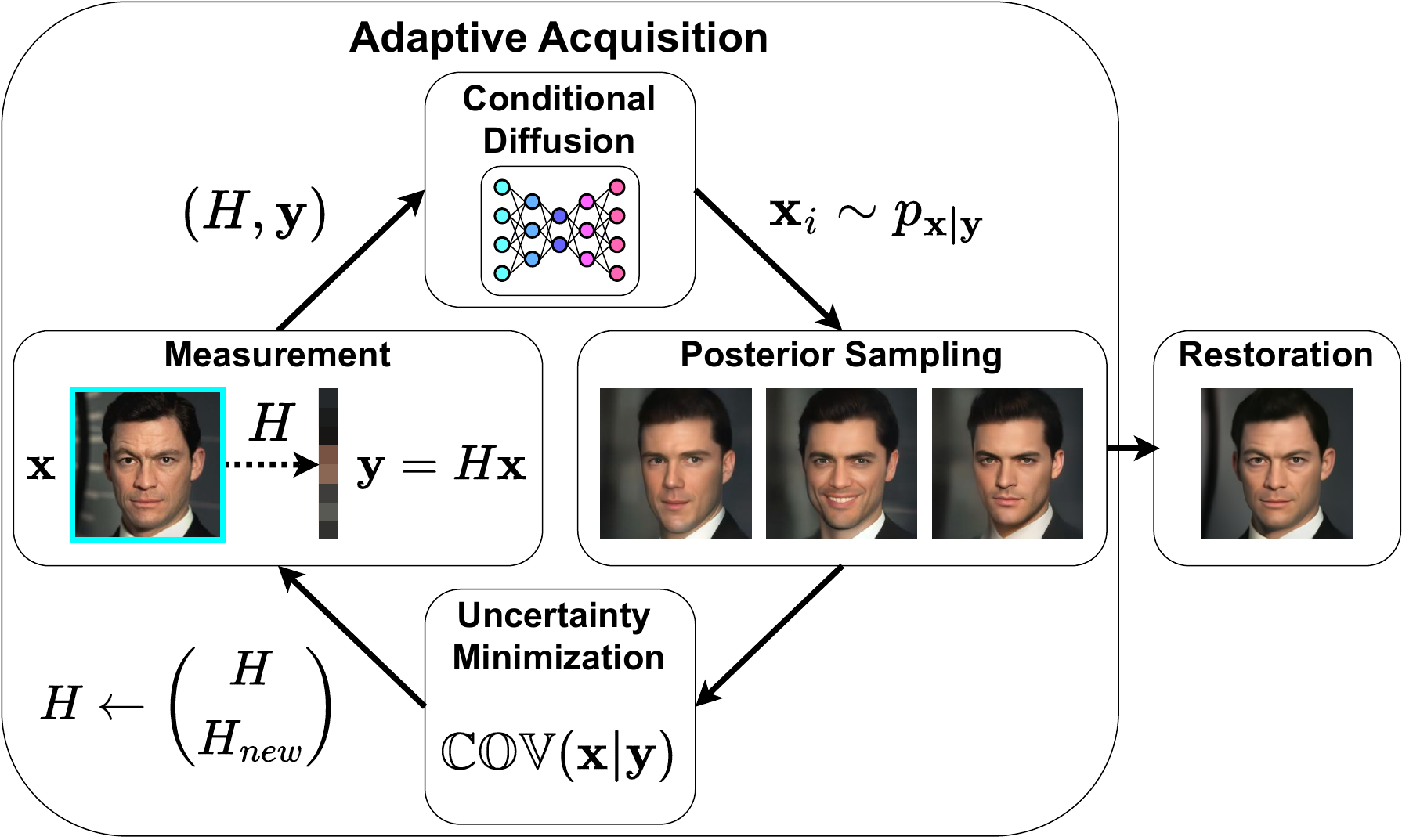}
  \caption{\textbf{A diagram of AdaSense.} In each acquisition step, the diffusion model generates conditional posterior samples, which are then used to estimate the posterior's covariance. The ground truth image (highlighted in blue) is measured using our newly-chosen sensing matrix, corresponding to the directions of highest uncertainty in the posterior distribution. This cycle continues until sufficient measurements are acquired. Finally, AdaSense leverages these measurements to restore the final image.
  }
  \label{fig:diagram}
\end{figure}


Existing methods for measurement selection are typically either heuristic or involve training complex models on simulated subsampled data. The complexity escalates when these methods are adaptive, as they necessitate fine-tuning the algorithm to the specific modality and subsampling schemes applied~\cite{van2021active, pineda2020active, bakker2020experimental}.
Consequently, such approaches are often limited in their modality, and struggle to adapt to changes within their domain.

Driven by recent advancements in zero-shot diffusion-based methods for solving linear inverse problems~\cite{ho2020denoising, sohl2015deep, song2020score, dhariwal2021diffusion, vahdat2021score, latent_diffusion}, we propose AdaSense, an image reconstruction technique for active measurement acquisition. These methods leverage pre-trained diffusion models~\cite{ho2020denoising, sohl2015deep, song2020score} to capture the data distribution, modifying the sampling algorithm to function as posterior distribution samplers. Using candidate posterior samples, AdaSense is able to quantify the reconstruction uncertainty associated with each possible future measurement, guiding the selection of the optimal subsequent measurement through a sequential algorithm. By employing zero-shot diffusion-based methods, AdaSense eliminates the need for additional training or fine-tuning, making it applicable across varied domains and acquisition schemes. Moreover, AdaSense operates under the assumption that access to training data is restricted, which is common in the sensitive medical domain. We offer several key insights to ensure AdaSense is computationally efficient.


To validate our approach, we apply AdaSense in various settings, including general image reconstruction and active acquisition of medical images. For general reconstruction, we compare AdaSense to other common subsampling techniques on facial images, and attempt to quantify the performance gain of adaptive restoration compared to non-adaptive alternatives. In the medical image domain, we showcase AdaSense as an active acquisition tool for both k-space magnetic resonance imaging (MRI) subsampling and sparse-view computed tomography (CT). In such scenarios, measurement subsampling may expedite the lengthy acquisition process and reduce patient exposure to radiation~\cite{Mullainathan2022, knoll2020fastmri, zbontar2018fastmri, zhang2019reducing, wang2023active}. To the best of our knowledge, AdaSense is the first active acquisition algorithm that both utilizes generative priors and is training-free, making it a unique and potentially impactful contribution to the field.

%

\begin{figure}[t]
  \centering
  \includegraphics[width=\textwidth]{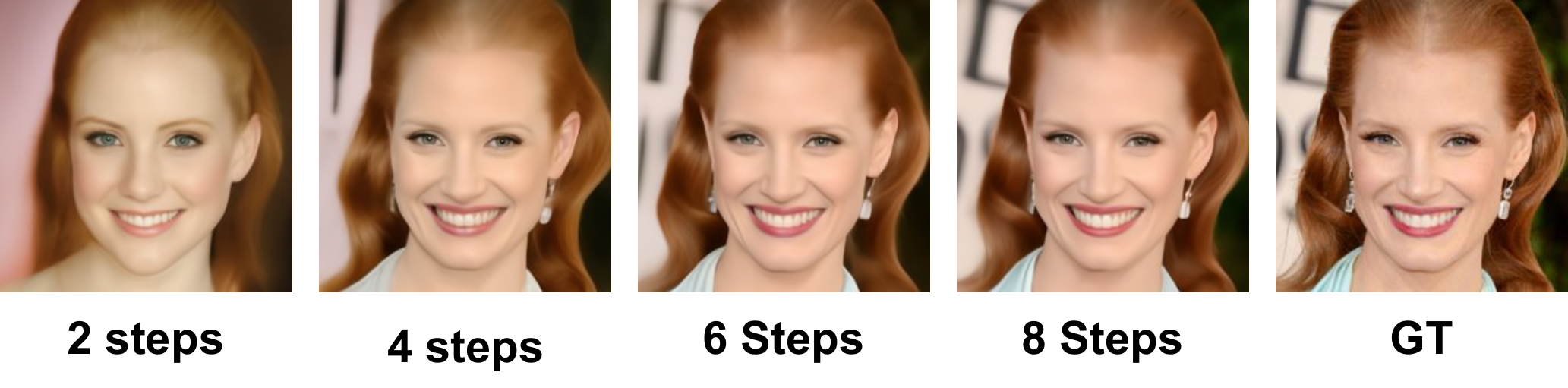}
  \caption{\textbf{Progressive restoration using AdaSense, demonstrating the increasing image reconstruction quality with the accumulation of additional measurements.} The number of reconstruction steps is denoted below each image, each step adds 24 measurements.}
  \label{fig:progression}
\end{figure}

\section{Related Work}

\paragraph{Diffusion Models.} Diffusion models have emerged as the leading approach for image generation and modeling~\cite{ho2020denoising, sohl2015deep, song2020score, dhariwal2021diffusion, vahdat2021score, latent_diffusion}. Diffusion models have been successfully extended to diverse domains, including audio and speech~\cite{kong2020diffwave, jeong2021diff, manor2024zeroshot}, video~\cite{luo2023videofusion, liu2024sora}, and medical data~\cite{song2023solving, chung2022score, kawar2023gsure, alcaraz2023diffusion}. Their capabilities extend beyond generation, excelling at solving inverse problem~\cite{saharia2021image, saharia2022palette, chung2023diffusion}, including many types of image reconstruction tasks. Recent years have seen a surge in zero-shot diffusion-based inverse problems solvers~\cite{choi2021ilvr, kawar2021snips, kawar2022denoising, song2023pseudoinverse, mardani2023variational, graikos2022diffusion, wang2022zero, chung2022improving} using pre-trained diffusion models. These methods modify the reverse diffusion process by leveraging the available measurements $\rvy$ through the exploitation of the known degradation form. AdaSense utilizes such a posterior sampler as its core, adapting to any chosen degradation or sensing matrix for measurement acquisition.

\paragraph{Compressed Sensing.} Compressed Sensing attempts to find an optimal sensing matrix for reconstruction of a given set of signals, initially employed using a sparsifying basis~\cite{donoho2006compressed, candes2006compressive}, and later using generative priors~\cite{bora2017compressed, kamath2019lower}.
Adaptive CS integrates reconstruction with the continual optimization of the sensing matrix~\cite{ji2008bayesian}. Similar works~\cite{seeger2011large, arias2012fundamental, haupt2012sequentially, malloy2014near, haupt2009adaptive, nakos2018improved} have refined the algorithmic underpinnings of adaptive compressed sensing, albeit relying on assumptions regarding data sparsity. In recent years, the integration of deep learning has propelled adaptive CS methods even further~\cite{haussmann2019deep, mi2020learning}, with some methods incorporating the deep priors of generative models\cite{van2021active, van2023active, zhang2018multi}. However, many existing methods are confined to specific domains and are thus tuned for particular tasks~\cite{zhang2019reducing, shen2020learning}. Additionally, certain methods are limited in the types of degradation they can handle, focusing solely on subsampling of pixels\cite{van2021active, van2023active}. In this work, we propose a novel perspective on the application of adaptive CS as a unified framework, using a stochastic posterior sampler. As these samplers grow in quantity and quality following the recent advances of deep learning and diffusion models, we hope to lay the foundation of a practical and straightforward adaptive CS algorithm.
\paragraph{Medical Imaging.} In medical imaging, such as MRI or CT, measurement subsampling may expedite the lengthy acquisition process and reduce patient exposure to radiation~\cite{zhang2019reducing, barkan2013adaptive, Mullainathan2022}. In MRI, images are obtained in the Fourier spectrum, named the k-space, using magnetic fields. Common MRI subsampling involves acquiring only a subset of possible k-space frequencies. Extensive research efforts have been directed towards developing reconstruction algorithms for subsampled data~\cite{wang2016accelerating, hammernik2018learning, schlemper2017deep, chen2018variable, zhu2018image, wang2020deepcomplexmri}. Specifically, diffusion models have demonstrated their prowess in the restoration domain for MR images~\cite{song2023solving, wu2023wavelet, jalal2021robust}, yet previous work falls short of exploring their potential for active MRI acquisition.

Current approaches for active MRI acquisition often involve deep neural network (DNN) training~\cite{jin2019self, van2021active, van2023active}, in many cases using reinforcement learning for policy selection~\cite{zhang2019reducing, pineda2020active, bakker2020experimental}. Despite impressive results, previous MRI active acquisition techniques' reliance on DNN training is both costly and a limiting factor when adapting to new subsampling schemes. In contrast, AdaSense only requires a pre-trained diffusion model. This allows the same algorithm to be applied seamlessly across diverse subsampling patterns, such as switching from vertical to horizontal frequency sampling in the MRI domain.

CT images are acquired using X-ray projections and reconstructed by means of tomography. Sparse-view CT aims to reduce acquisition time and radiation dose by utilizing fewer projection angles for reconstruction. Recent work has introduced deep learning, including diffusion and score-based models, for sparse-view CT reconstruction~\cite{wu2023wavelet, wang2023review, guan2023generative}. Yet, in these methods the subsampling scheme is assumed to be pre-determined and non-adaptive. Similar to MRI, the active acquisition of sparse-view CT has been explored using reinforcement learning for DNN training~\cite{shen2020learning, wang2023sequential}. In our work, we show AdaSense offers a training-free approach, demonstrating its potential as an active sparse-view acquisition tool for CT reconstruction.

\paragraph{Blind Inverse Problems.} Recent research in blind image restoration utilizes diffusion-based inverse problem solvers to adapt the sensing matrix~\cite{murata2023gibbsddrm, chung2023parallel, laroche2024fast}. While our approach shares similarities with these methods by using diffusion-based approaches for determining the sensing matrix, our goals differ fundamentally. We focus on actively selecting the optimal series of measurements taken from an undamaged image, rather than attempting to revert an unknown degradation in an already captured image.

\section{Method}
\label{sec:method}
We are interested in reconstructing a signal $\rvx\in\R^D$ from $d<D$ linear measurements of the form
\begin{equation}
\label{eq:lin-inv}
    \rvy = \mH\rvx,
\end{equation}
where $\mH\in\R^{d\times D}$ is the sensing matrix and $\rvy\in\R^d$ is the vector of measurements. We assume that $\rvx$ is a random vector whose distribution $p_{\rvx}$ is either known or has been learned by a generative model based on a large corpus of training samples. 
Our goal is to determine the matrix $\mH$ that allows the best possible recovery of $\rvx$ in the mean squared error (MSE) sense. Importantly, we assume we may select the measurements (\ie the rows of $\mH$) in a sequential manner, taking advantage of information gained from previous measurements.

Let us first discuss the simplified non-adaptive setting, in which the entire matrix $\mH$ is determined a-priori without dependence on the measurements. In this setting, the optimization problem we would like to solve is
\begin{equation}
\label{eq:optim-rec}
    \argmin_{\mH, f} \E \left[ \| f(\mH\rvx) - \rvx \|^2 \right],
\end{equation}
where $f:\R^{d}\to\R^{D}$ is a function that reconstructs $\rvx$ from the measurements $\mH \rvx$. In principle, the optimal $\mH$ and $f(\cdot)$ may be determined by training an auto-encoder with a linear encoder $\mH$ and a nonlinear decoder $f(\cdot)$. However, as will become clear below, adapting this strategy to the adaptive CS setting is computationally impractical. 
A more scalable approach is to restrict the decoder function to be an affine transformation of the form $f(\rvy) = \mU\rvy + \rvb$, relying on the assumption that the optimal $\mH$ for this case is also nearly optimal for \cref{eq:optim-rec}. In this setting, the optimization problem becomes
\begin{equation}
\label{eq:pca}
    \argmin_{\mU, \rvb, \mH} \E \left[ \| \mU \mH \rvx + \rvb - \rvx \|^2 \right].
\end{equation}
This problem corresponds to principal component analysis (PCA)~\cite{pca, hotelling1933analysis}. Therefore, the optimal sensing matrix $\mH$ has the top $d$ eigenvectors of $\Cov[\rvx]$ as its rows, the optimal decoding matrix is $\mU=\mH^\top$, and the optimal bias is $\rvb = (\mI - \mH^\top \mH)\E[\rvx]$. 
Namely, PCA minimizes the \emph{uncertainty}, defined as the MSE attained by the linear minimum-MSE (MMSE) predictor of $\rvx$ based on the measurements $\rvy$. This is achieved by choosing the measurement matrix $\mH$ accordingly to minimize this error.


We next address the \emph{adaptive} case, where the rows of $\mH$ are chosen sequentially, such that each row (or chunk of rows) is chosen based on the measurements acquired with the preceding rows. We take a greedy approach, where in each stage we choose the new rows so as to allow minimization of the reconstruction error at that stage, without taking into consideration their effect on the reconstruction error at future stages. 
We denote by $\mH_{i:j}$ the sub-matrix of $\mH$ corresponding to rows $i$ to $j$ (including row $i$ and excluding row $j$) 
and denote by $\rvy_{i:j}$ the corresponding measurements, $\rvy_{i:j} = \mH_{i:j}\rvx$. We aim to construct the whole matrix $\mH$ in $N$ steps, selecting $r$ new rows in each step. Therefore, at step~$n$, we rely on the previous $nr$ measurements, $\rvy_{0:nr}$. 
Following the same PCA strategy discussed for the non-adaptive case, we propose choosing the sub-matrix $\mH_{nr:nr+r}$ at stage $n$, as 
\begin{align}
\label{eq:best-cond-deg}
        \nonumber 
        \mH_{nr:nr+r}&=\argmin_{\mV,\mU,\rvb,\tilde{\mH}} \E \left[\left\| \begin{pmatrix}\mV & \mU \end{pmatrix}\begin{pmatrix}
             \mH_{0:nr}  \\
             \tilde{\mH}
        \end{pmatrix} \rvx + \rvb - \rvx  \right\|^2 \middle|\rvy_{0:nr}\right]  \\
        &=\argmin_{\mV,\mU,\rvb,\tilde{\mH}} \E \left[ \left\Vert  \mV \rvy_{0:nr} + \mU \tilde{\mH} \rvx + \rvb - \rvx \right\Vert^2  \middle| \rvy_{0:nr}\right] .
\end{align}
Here, $ \begin{pmatrix}\mV & \mU \end{pmatrix} $ serves as the linear reconstruction matrix and we used the fact that $\mH_{0:nr}\rvx=\rvy_{0:nr}$. 
Due to the conditioning on $\rvy_{0:nr}$, the vector $\mV\rvy_{0:nr} + \rvb$ can be viewed as a deterministic bias term (playing the role of $\rvb$ in \cref{eq:pca}). Therefore, similarly to \cref{eq:pca}, the optimal $\tilde{\mH}$ in this case corresponds to the top $r$ eigenvectors $\Cov[\rvx|\rvy_{0:nr}]$.

Once $\mH_{nr:nr+r}$ is determined, we can use it to obtain $r$ new measurements, $\rvy_{nr:nr+r}=\mH_{nr:nr+r}\rvx$, and append them to our previous ones as $\rvy_{0:nr+r}=\begin{pmatrix} \rvy_{0:nr}^\top & \rvy_{nr:nr+r}^\top \end{pmatrix}^\top$. We may repeat this process until we have chosen all $N r$ rows of $\mH$. Finally, we may restore our final measurements $\rvy$ with a nonlinear reconstruction function $f(\rvy)$.

The iterative approach described above requires knowing the posterior covariance matrix $\Cov[\rvx|\rvy_{0:nr}]$ in each stage, or at least its top eigenvectors. 
To obtain an approximation of these principal components, we propose harnessing zero-shot posterior sampling methods that are based on a pre-trained diffusion model~\cite{kawar2021snips, chung2023diffusion, kawar2022denoising, song2023pseudoinverse, mardani2023variational, graikos2022diffusion, wang2022zero, chung2022improving}. These methods allow generating samples from the posterior distribution of $\rvx$ given $\rvy_{0:nr}$ for problems in which $\rvy_{0:nr}$ is a linear transformation of $\rvx$, as in our setting. Given a set of $s$ such posterior samples, we can apply PCA to determine the top eigenvectors of the (empirical) posterior covariance. Our sampling algorithm, Adasense, is presented in \cref{alg:AdaSense}, including the constrained measurement scenario which we describe in \cref{sec:constrained} below. A diagram of our proposed method may be found in \cref{fig:diagram}, as well as a qualitative example of progressively increasing restoration quality with the accumulation of additional measurements in \cref{fig:progression}.

\begin{algorithm}[t]
\caption{AdaSense}
\label{alg:AdaSense}
 \begin{algorithmic}
    \Require Generative model for $p_\rvx$, number of steps $N$, number of measurements per step $r$, number of posterior samples $s$.
    \State \textbf{initialize} $\rvy_{0:0}$ as an empty vector of measurements.
    \For{$n\in\{0 : N-1\}$}
        \State $\{\rvx_i\}_{i=1}^s \sim p_{\rvx|\rvy_{0:nr}}$  \Comment{\textcolor{gray}{generate $s$ posterior samples}}
        \State $\{\rvx_i\}_{i=1}^s \gets \{\rvx_i - \frac{1}{s} \sum_{j=1}^s \rvx_j \}_{i=1}^s$
        \If{not Constrained Measurements} \Comment{\textcolor{gray}{select $r$ principal components}}
            \State $\mH_{nr:nr+r} \gets \text{top $r$ right singular vectors of }\begin{pmatrix}\rvx_1 & \dots & \rvx_s \end{pmatrix}^\top $  
        \Else  \Comment{\textcolor{gray}{select $r$ best sensing rows}}
            \State $\mH_{nr:nr+r} \gets \argmax_{\tilde{\mH}\in\mathcal{H}} \sum_{i=1}^s \rvx_i^\top \tilde{\mH}^\dagger \tilde{\mH} \rvx_i $
        \EndIf
        \State $\rvy_{0:nr+r} \gets \text{Append} \left( \rvy_{0:nr}, \mH_{nr:nr+r} \rvx \right)$ \Comment{\textcolor{gray}{measure the real image $\rvx$}}
    \EndFor
    \State \textbf{return} $\rvx_1 = f(\rvy_{0:Nr}) $ \Comment{\textcolor{gray}{posterior sampling or alternative restoration}}
 \end{algorithmic}
\end{algorithm}

\subsection{Constrained Measurement}
\label{sec:constrained}
In certain real-world applications, the sensing matrix is constrained due to physical limitations. One notable example is MRI, in which the signal is measured in $k$-space, so that the rows of $\mH$ are restricted to be rows of the Fourier transform matrix. Another important example is CT, in which the signal is measured in Radon space. In this setting, the rows of $\mH$ are constrained to be rows of the Radon transform matrix. To adapt our method to such scenarios, we therefore need to choose the best option from a predefined set $\mathcal{H}$ of feasible sensing matrices. In other words, we need to solve \cref{eq:best-cond-deg} under the constraint $\tilde{\mH}\in\mathcal{H}$. As we show in App.~\ref{app:proof-of-optimal-constrained}, the matrix $\tilde{\mH}\in\mathcal{H}$ achieving the optimum of this constrained optimization problem is the solution to
\begin{align}
\label{eq:optimal-constrained} 
        \argmax_{\tilde{\mH}\in\mathcal{H}}  \trace\left\{  \tilde{\mH} \Cov[\rvx|\rvy_{0:nr}]^2 \tilde{\mH}^\top \left( \tilde{\mH} \Cov[\rvx|\rvy_{0:nr}] \tilde{\mH}^\top\right)^{-1} \right\}.
\end{align}
Note that as opposed to the unconstrained case, if $\mathcal{H}$ is not the entire space $\R^{r\times D}$, then the solution to this optimization problem is not necessarily the top eigenvectors of  $ \Cov[\rvx|\rvy_{0:nr}]$. 

Unfortunately, in many practical cases the number of rows $r$ selected in each step is unavoidably large. For example, in MRI, we would often like to measure a whole column of the 2d Fourier transform  of the image at a time. Similarly in CT, we would often like to measure a whole column of the sinogram at a time. In such scenarios, attempting to directly approximate the solution to \cref{eq:optimal-constrained} becomes prohibitively expensive. This is because for the $r\times r$ matrix $\tilde{\mH} \Cov[\rvx|\rvy_{0:nr}] \tilde{\mH}^\top$ to be invertible, the rank of our approximation of the covariance  $\Cov[\rvx|\rvy_{0:nr}]$ must be at least $r$. This requires generating $s\geq r$ posterior samples in each step, which is impractical with today's zero-shot posterior samplers for the typical values of $r$ in \eg MRI and CT. To overcome this computational difficulty, we propose a sub-optimal solution, which nonetheless works quite well in practice. Specifically, rather than optimizing over  the reconstruction matrix $\mU$ in \cref{eq:best-cond-deg}, we set it to be $\tilde{\mH}^\dagger$, which is its optimal values in the unconstrained scenario. As we demonstrate in App.~\ref{app:proof-of-constrained}, this simplifies the problem to 
\begin{align}
    \label{eq:constrained} 
    \argmax_{\tilde{\mH}\in\mathcal{H}}\E \left[ (\rvx-\E[\rvx|\rvy_{0:i}])^\top \tilde{\mH}^\dagger \tilde{\mH} (\rvx-\E[\rvx|\rvy_{0:i}]) \middle|\rvy_{0:i} \right]. 
\end{align}
The solution to this problem can again be approximated using posterior samples, by replacing expectations by averages and exhaustively scanning all matrices in $\mathcal{H}$. This search space becomes impractically large to scan if we attempt to select several frequency columns in MRI or several sinogram columns in CT at once. In such cases, we use a heuristic, as detailed in App.~\ref{app:condiserations-constrained}. 


\subsection{Restoration}
\label{sec:versions}

Once we have obtained the optimal sensing matrix $\mH$ using the method outlined above, we seek a  function $f(\cdot)$ to restore our final set of measurements $\rvy_{0:Nr}$. The most straightforward approach would be to use the same zero-shot posterior sampler for the final restoration, generating a single or multiple reconstructions. 
Another alternative, involves using the average of several posterior sampler outputs. This average approximates the posterior mean, which is the minimum MSE (MMSE) estimator. 
A third alternative is to take the restoration function $f(\cdot)$ to be a neural network that is specifically tailored to the modality. ~\Cref{sec:mri} experiments with classical and deep learning reconstruction approaches specifically designed for MRI for improved restoration.

\begin{figure}[t]
  \centering
  \includegraphics[width=\textwidth]{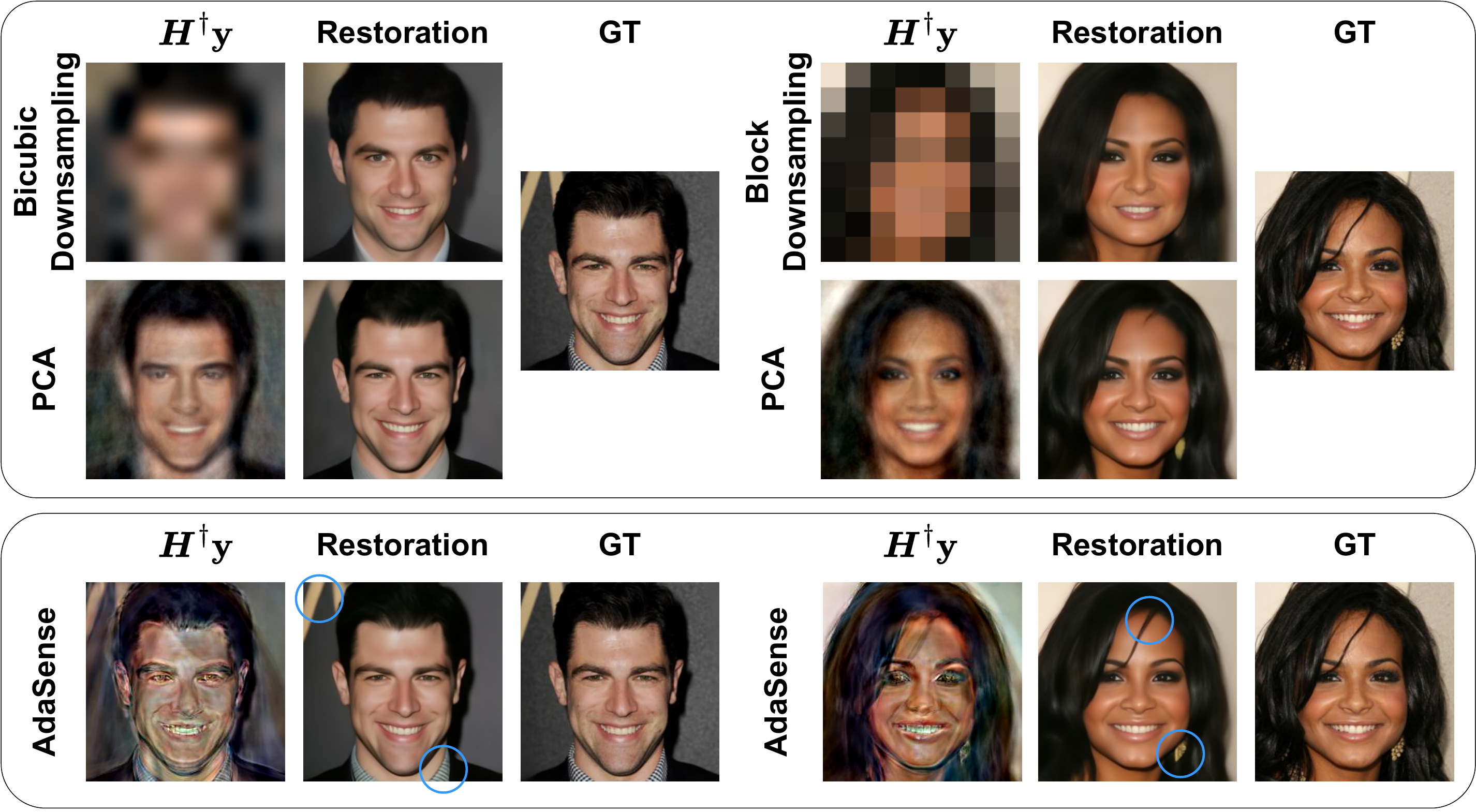}
  \caption{\textbf{Face image restoration using AdaSense.} 
    We illustrate how our adaptive approach (bottom) is better for image reconstruction than block-downsampling or bicubic-downsampling (top), where measurements have 192 elements. Notably, AdaSense successfully preserves some finer details (circled in blue). Also, we compare AdaSense with the optimal non-adaptive approach, using PCA (\cref{eq:pca}) on the distribution of training images. The images are restored using DDRM. 
    }
  \label{fig:celebahq-restoration}
\end{figure}

\subsection{Acceleration of Posterior Sampling}
\label{sec:acceleration}

AdaSense, as outlined in \cref{alg:AdaSense}, may be used with any posterior sampler. However, to avoid training for specific degradations we propose using a zero-shot diffusion restoration method. While the mechanics of different methods differ, many approaches solely rely on the computation of the Moore–Penrose pseudo-inverse $\mH^\dagger$ of the degradation matrix $\mH$. Also, several zero-shot diffusion restoration methods are consistent with the measurements $\rvy$, i.e. for any generated $\rvx$ the equation $\rvy = \mH\rvx$ holds\footnote{While consistency should hold by the definition of posterior sampling, not all zero-shot diffusion restoration methods guarantee consistency~\cite{chung2023diffusion, kawar2022denoising, song2023pseudoinverse}.}. Below, we describe how we can greatly increase the efficiency of our implementation using such approaches.

The computation of $\mH^\dagger$, could generally be computed for any degradation $\mH$ using a computationally expensive SVD. Because the degradation matrix $\mH$ is chosen at runtime in AdaSense, the SVD computation could potentially lead to slower sampling. Nevertheless, we offer several insights that explain why this repeated computation of SVD can be disregarded. Because a consistent posterior sampler is used, the variance along previously selected sensing matrix is necessarily zero. Therefore, for general measurements and cases of constrained measurements where the measurements are inherently non-overlapping (such as MRI subsampling), our choices for the next measurement $\tilde{\mH}$ are necessarily orthogonal to all previous measurements. This leads to the useful property where the low-rank SVD of our matrices $\mH_{0:nr}$ is $(U, \Sigma, V^\top) = (I, I, \mH_{0:nr})$, eliminating the need for additional computation. Further details on are provided in the supplementary material. In other cases, such as sparse-view CT reconstruction, limiting AdaSense to a small number of total measurements ensures that the matrix $\mH_{0:nr}$ remains low-rank, making SVD computation inexpensive.

\section{Experiments}

\begin{table}[t]
  \caption{\textbf{Comparisons of several DDRM-based restorations on different measurements of rank 192.} Also, the table depicts how changing the final restoration algorithm can decrease distortion at the expense of image quality, for the same chosen measurements.
  }
    \label{tab:celeba-table}
  \centering
  \begin{tabular}{@{}lcccc@{}}
    \toprule
    CELEBA-HQ 256 & PSNR$\uparrow$ & SSIM$\uparrow$ & LPIPS$\downarrow$ & DeepFace CosSim$\downarrow$ \\
    \midrule
    Block Downsampling    & 20.50 & 0.6128 & 0.3035 & 0.6134  \\
    Bicubic Downsampling & 20.86 & 0.6193 & 0.2964 & 0.5865 \\
    Walsh-Hadamard CS &  10.75 & 0.3874 & 0.5920 & 0.8836  \\
    Missing Pixels & 10.85 & 0.3865 & 0.6064 & 0.8782 \\
    PCA &  24.60 & 0.7190 & 0.2307 & 0.3010  \\
    AdaSense    & \textbf{26.20} & \textbf{0.7515} & \textbf{0.1950} &  \textbf{0.2674} \\
    \midrule
    AdaSense (Mean)   & \textbf{26.87} & \textbf{0.7713} & 0.2449 & \textbf{0.2614} \\
    \bottomrule
  \end{tabular}
\end{table}

\subsection{General Images}
\label{sec:general-images}
We begin by testing AdaSense on $256\times256$ face images taken from the CelebA-HQ~\cite{liu2015celeba, karras2017progressive} validation dataset. 
The pre-trained model from SDEdit~\cite{meng2021sdedit} was used with DDRM~\cite{kawar2022denoising} for diffusion-based sampling. 
We evaluate AdaSense's capabilities to select the best possible measurements by comparing reconstructions made using AdaSense's proposed sensing matrix to reconstructions from different common degradations. All degraded measurements are of dimension 192 and have been restored using the same algorithm (DDRM~\cite{kawar2022denoising}).
AdaSense employed $N=8$ consecutive iterations, selecting $r=24$ elements per iteration. \Cref{fig:celebahq-restoration} showcases a qualitative comparison of AdaSense with other restoration approaches, highlighting its superior ability to preserve fine details and subject identity compared to alternatives.
Qualitative result in \cref{tab:celeba-table} reveal that AdaSense outperforms other non-adaptive methods across all metrics, including PSNR, SSIM, LPIPS~\cite{lpips}, and DeepFace~\cite{serengil2020lightface} cosine similarity.  Notably, AdaSense surpasses a non-adaptive PCA approach that selects the optimal sensing matrix following \cref{eq:pca} using real training images. This achievement highlights the effectiveness of AdaSense, especially considering it relies solely on approximated generated samples. In addition, we show how using the mean of the posterior sampler as suggested in \cref{sec:versions} lowers the distortion of the final reconstruction (measured in PSNR and SSIM) while sacrificing the perceptual quality (measured in LPIPS).

\begin{wrapfigure}{R}{0.5\textwidth}
  \centering
  \includegraphics[width=0.5\textwidth]{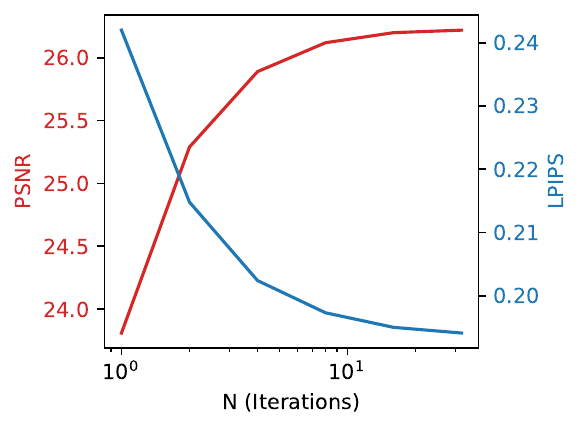}
  \caption{\textbf{Graphs of the effect of adaptivity on restoration.} AdaSense is used to restore the same set of images using the same total number of measurements $N\cdot r$, but with different numbers of iterations $N$ and measurements per iteration $r$. The image restoration improves considerably the more adaptive the algorithm.} 
  \label{fig:adaptivity}
\end{wrapfigure}

\subsection{Importance of Adaptivity}
\label{sec:adaptivity}
The hyperparameter $N$ governs the number of iterations in the measurement acquisition process. Within a constant total number of measurements $N \cdot r$, AdaSense is deemed `less adaptive' when $r$ is larger at the expense of $N$, and `more adaptive' otherwise. For $N=1$, a single iteration is used, essentially eliminating the adaptive component of the algorithm.


To validate the significance of adaptivity in the restoration process, we compare our method's performance using different numbers of iterations for measurement acquisition, while maintaining the total number of measurements $N\cdot r$. This ensures a fair comparison focused solely on the effect of adaptivity. Also, the value of $s$ remains consistently proportional to $r$ at a ratio of $\nicefrac{4}{3}$, ensuring consistent computational demands (measured by the total number of generated samples) across experiments. We retain the posterior sampler from \cref{sec:general-images}, and adjust the AdaSense hyperparameters $N, r, s$ to change the adaptivity of our algorithm. The results, illustrated in \cref{fig:adaptivity} demonstrates that both PSNR and LPIPS~\cite{lpips} improve as the number of iterations used for measurement acquisition increases. This trend underscores the positive impact of adaptivity on the system's robustness and effectiveness, highlighting its crucial role within our proposed method.

\begin{table}[t]
  \caption{\textbf{Comparison of AdaSense with non-adaptive MRI restoration algorithms.} The table compares subsampling masks generated according to~\cite{bridson2007fast, zbontar2018fastmri, knoll2020fastmri} with sensing matrices selected by Adasense. The number following R denotes the acceleration factor. The table displays results for three reconstruction methods: DDRM~\cite{kawar2022denoising}, mean of 16 DDRM samples, and $L1$ wavelet regularized reconstruction~\cite{lustig2007sparse}.}
    \label{tab:mri-table}
  \centering
  \begin{tabular}{@{}lcccccc@{}}
    \toprule
    & \multicolumn{2}{c}{\textbf{DDRM}} & \multicolumn{2}{c}{\textbf{DDRM Mean}} & \multicolumn{2}{c}{\textbf{Wavelet}} \\
    \textbf{Vertical Subsampling Mask} \quad & PSNR$\uparrow$ & SSIM$\uparrow$ & PSNR$\uparrow$ & SSIM$\uparrow$ & PSNR$\uparrow$ & SSIM$\uparrow$ \\
    \midrule

    Random R10    & 24.56 & 0.4786 & 25.29 & 0.5067 & 22.50 & 0.4365 \\
    Equi-spaced R10    & 24.73 & 0.4767 & 25.44 & 0.5043 & 22.56 & 0.4339 \\
    AdaSense R10 & \textbf{27.01} & \textbf{0.5229} & \textbf{27.58} & \textbf{0.5431} & \textbf{25.18} & \textbf{0.4874} \\
    \midrule
    \textbf{General Subsampling Mask} \quad & PSNR$\uparrow$ & SSIM$\uparrow$ & PSNR$\uparrow$ & SSIM$\uparrow$ & PSNR$\uparrow$ & SSIM$\uparrow$ \\
    \midrule
    Poisson Disk R400    & 23.06 & 0.3728 & 23.86 & 0.4010 & 18.22 & 0.3504 \\
    AdaSense R400 & \textbf{25.26} & \textbf{0.4124} & \textbf{25.76} & \textbf{0.4314} & \textbf{23.32} & \textbf{0.4010} \\
    \midrule
    Poisson Disk R200    & 24.31 & 0.3999 & 24.95 & 0.4234 & 22.50 & 0.3619 \\
    AdaSense R200 & \textbf{25.94} & \textbf{0.4331} & \textbf{26.39} & \textbf{0.4510} & \textbf{24.44} & \textbf{0.4219} \\
    \bottomrule
  \end{tabular}
\end{table}

\begin{table}[!t]
\centering
\caption{\textbf{Comparisons of MRI Active Acquisition Algorithms.} This table compares the performance of AdaSense with other MRI active acquisition algorithms. The acceleration factors are denoted by R, and the number of pre-set low-frequencies in L, following~\cite{pineda2020active}. We also include results for the non-adaptive baseline strategy `Low-to-High'~\cite{pineda2020active}, and the `Greedy Oracle'~\cite{pineda2020active} approach, a theoretical upper-bound on the achievable performance. As shown in the table, AdaSense demonstrates comparable performance  to training-based methods.}
    \label{tab:mri-active-comparison}
  \centering
  \begin{tabular}{@{}lcccc@{}}
    \toprule
     & \multicolumn{2}{c}{\textbf{R8 - 30L}} & \multicolumn{2}{c}{\textbf{R16 - 2L}} \\
    \textbf{Method} & PSNR$\uparrow$ & SSIM$\uparrow$ & PSNR$\uparrow$ & SSIM$\uparrow$ \\
    \midrule
    Low-to-High (non-adaptive)~\cite{pineda2020active}      & 28.32 & 0.5948 & 25.66 & 0.4908 \\
    Reducing Uncertainty~\cite{zhang2019reducing}           & 28.66 & 0.6021 & 26.43 & 0.5386\\
    SS-DDQN~\cite{pineda2020active}                         & 28.99 & 0.6129 & 27.70 & 0.5691 \\
    DS-DDQN~\cite{pineda2020active}                         & 28.98 & 0.6135 & 27.60 & 0.5650\\
    Greedy Oracle~\cite{pineda2020active}                   & 29.22 & 0.6264 & 28.23 & 0.5846\\
    AdaSense (Reconstructor)                                       & 28.89 & 0.6108 & 27.51 & 0.5547\\
    
    \bottomrule
  \end{tabular}
\end{table}




\begin{figure}[t]
  \centering
  \includegraphics[width=\textwidth]{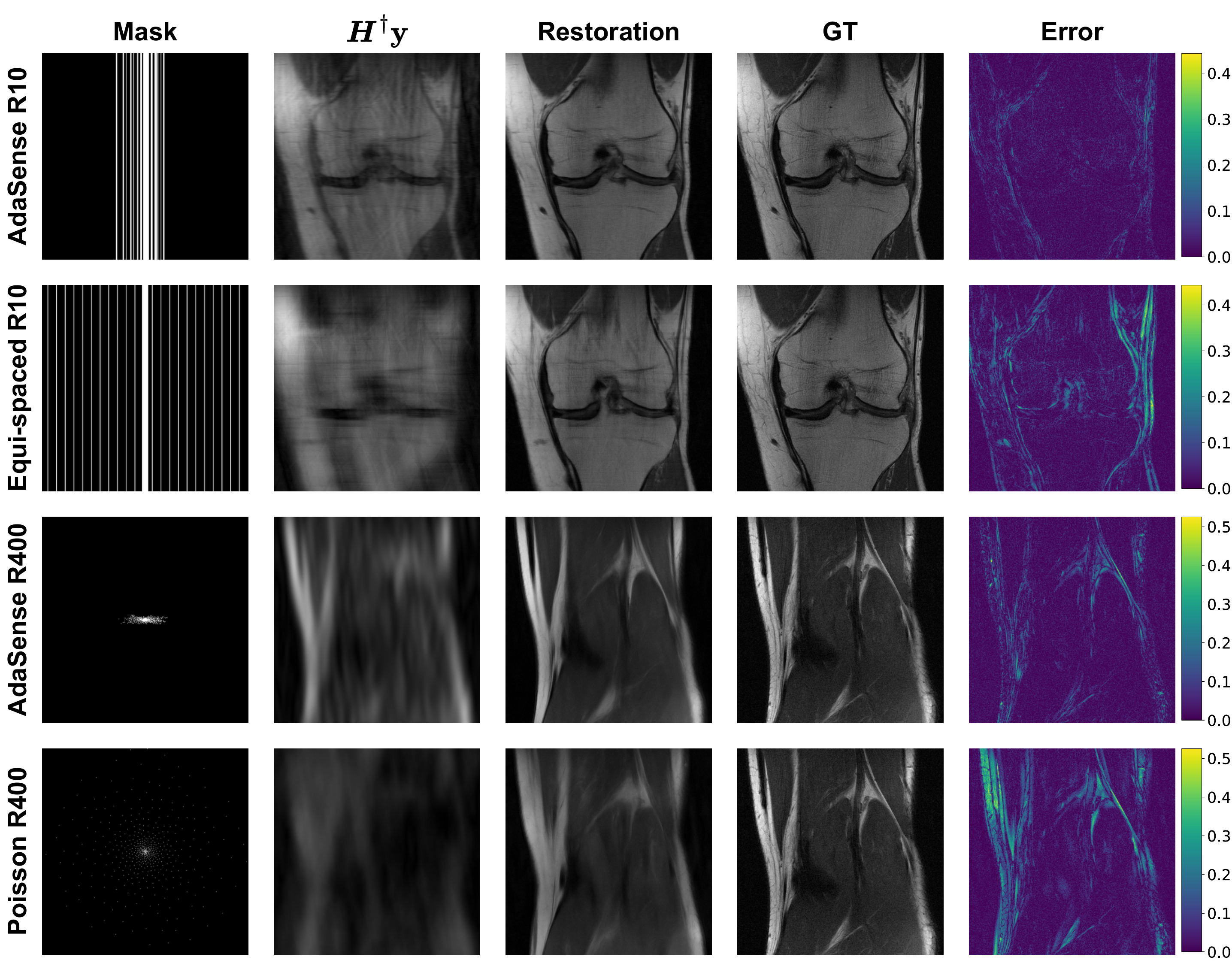}
  \caption{\textbf{MRI active restoration using AdaSense.} We compare various subsampling masks with AdaSense, restoring vertical subsampling with acceleration 10 and subsampling with acceleration 400. Images are reconstructed using DDRM~\cite{kawar2022denoising}. Each column of the figure shows, from left to right: the subsampling mask, the zero-filled measurements, the restored image, the ground-truth image, and the absolute error. The central 320×320 region is cropped and the image intensity is displayed.}
    \label{fig:mri-restoration}
\end{figure}

\begin{figure}[t]
  \centering
  \includegraphics[width=\textwidth]{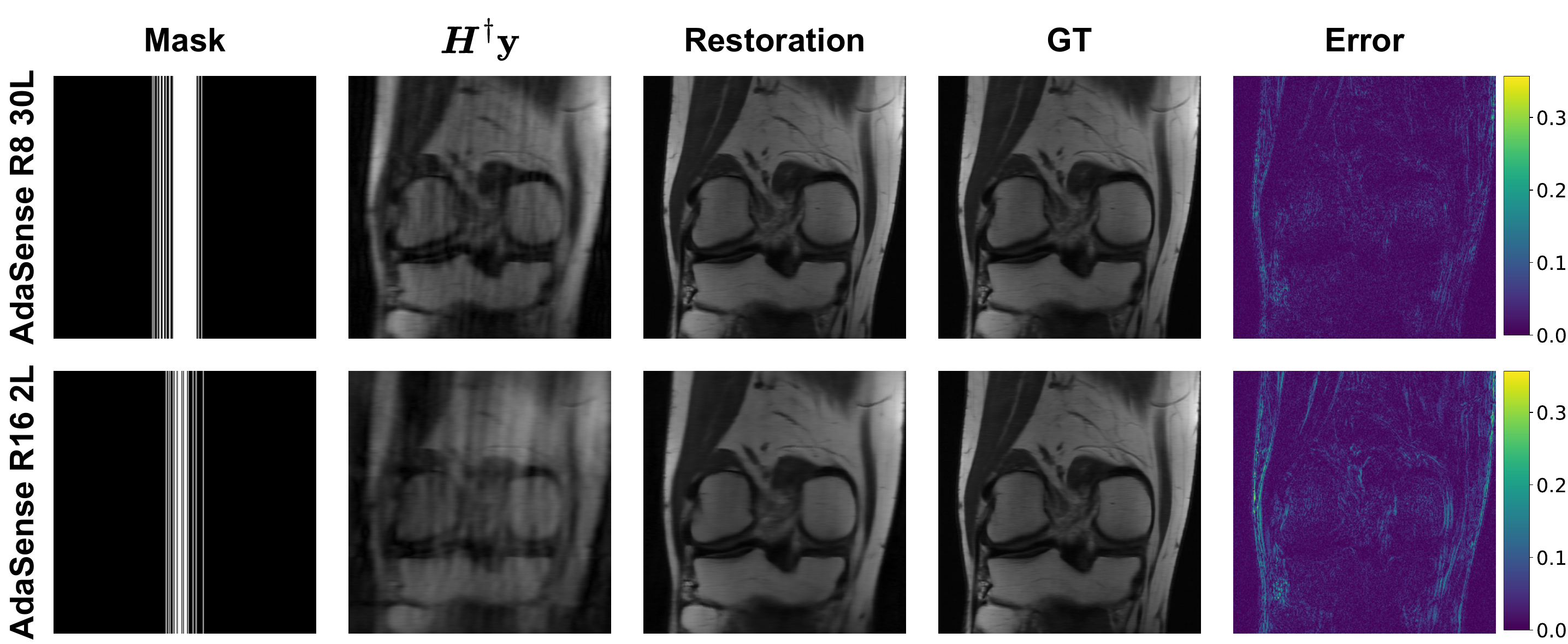}
  \caption{\textbf{MRI active restoration using AdaSense, using reconstructor models from~\cite{pineda2020active}.} The images depict image intensity for the central $320\times320$ region, applying vertical acceleration factors 8 and 16, following the strategies 30L and 2L respectively. Each column displays, from left to right: the subsampling mask, the zero-filled measurements, our AdaSense reconstruction, the ground-truth image, and the absolute error.}
    \label{fig:mri-act-restoration}
\end{figure}

\subsection{Active MRI Subsampling}
\label{sec:mri}
In this section, we demonstrate the usefulness of AdaSense in active MRI subsampling. 
We use a diffusion model for complex $640\times368$ single-coil knee MRI images which we train on the FastMRI~\cite{zbontar2018fastmri, knoll2020fastmri} dataset, 
following established data pre-processing~\cite{pineda2020active} and model architecture and training procedures~\cite{ho2020denoising}.
We apply AdaSense for image reconstruction using DDRM~\cite{kawar2022denoising} under various acceleration schemes. \Cref{tab:mri-table} compares AdaSense to non-adaptive subsampling masks, including general subsampling with acceleration factor of $200$ and $400$ and vertical subsampling with acceleration factor of $10$, created following~\cite{bridson2007fast} and~\cite{zbontar2018fastmri, knoll2020fastmri} respectively. 
Images are restored using posterior sampling or with a classical wavelet approach~\cite{lustig2007sparse}.
We measure PSNR and SSIM on the central $320\times320$ region of the reconstructed MR images, aligning with the evaluation method used in similar works~\cite{zbontar2018fastmri, pineda2020active}. As shown, AdaSense outperforms all non-adaptive measurements. Visualizations of the reconstructions are provided in \cref{fig:mri-restoration}. Furthermore, \cref{tab:mri-active-comparison} compares AdaSense against established  MRI active acquisition methods~\cite{zhang2019reducing, pineda2020active}. For a fair comparison, all approaches use the same final reconstruction model employed in the referenced works, termed `Reconstructor'. We include results for the non-adaptive baseline strategy `Low-to-High'~\cite{pineda2020active}, and the theoretical upper-bound `Greedy Oracle'~\cite{pineda2020active}. Specifically, the `Greedy Oracle' upper-bound is computed by searching all possible future measurements during each acquisition step, and selecting the ones minimizing the reconstruction error relative to the \emph{ground truth}, which is unavailable in real settings. Consistent with~\cite{pineda2020active}, frequency selection utilizes only the image intensity of the central $320\times320$ region, where model evaluation takes place (details in App.~\ref{app:condiserations-constrained}). Visual examples are included in \cref{fig:mri-act-restoration}. AdaSense remains competitive with similar training-based methods, despite requiring no additional training beyond the pre-trained diffusion model, and never encountering degraded data during the diffusion model's training. This underscores AdaSense's advantage in adapting to any subsampling scheme. 

\subsection{Sparse-View CT Reconstruction}
\begin{figure}[t]
  \centering
  \includegraphics[width=\textwidth]{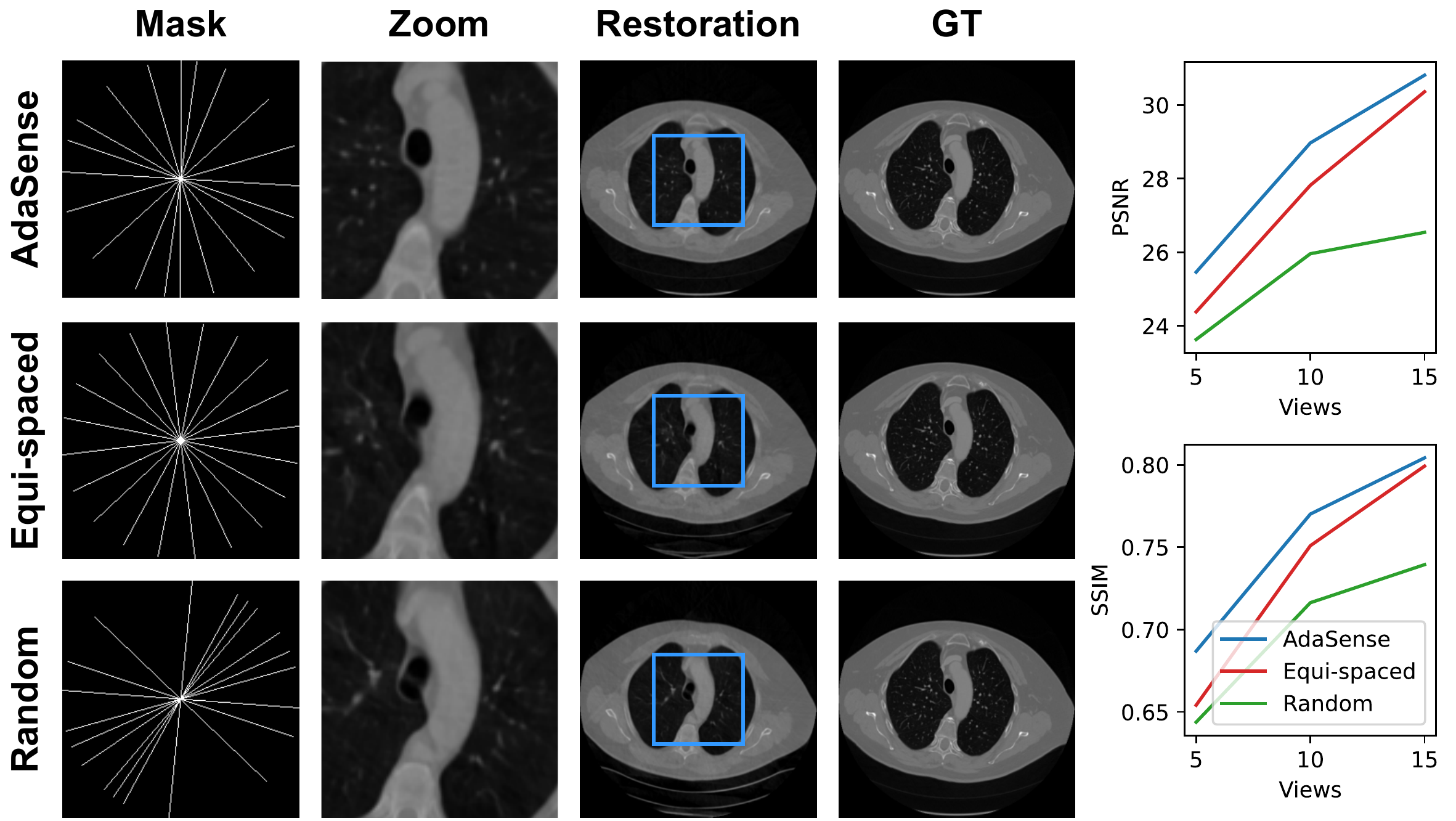}
  \caption{\textbf{CT sparse-view image restoration using AdaSense.} 
    Left: Reconstructions obtained using various acquisition strategies. Each reconstruction is accompanied by its corresponding mask, which illustrates the selected acquisition angles. Additionally, an enlarged center crop of the respective reconstruction is provided for detailed visualization. Right: Quantitative evaluations of CT restorations using PSNR and SSIM. Both these qualitative and quantitative findings results demonstrate the potential of AdaSense for active acquisition in diverse domains.}
  \label{fig:ct-restoration}
\end{figure}

AdaSense can also be used for sparse-view CT reconstruction, by selecting the best projection angles for restoration. For this experiment, we have trained a simple diffusion model following~\cite{ho2020denoising} for $256\times256$ CT images from on the DeepLesion\cite{yan2018deeplesion} dataset. We simulate restoration of parallel-beam projections and measure PSNR and SSIM only only within the valid region of the simulated projections. Similar to previous experiments, we use DDRM~\cite{kawar2022denoising} as our posterior sampler. Applying AdaSense to various acceleration schemes on samples from the validation set yields promising results, as showcased in~\cref{fig:ct-restoration} through both qualitative and quantitative data for sparse-view acquisitions.

\section{limitations}

While AdaSenseAdaSense demonstrates impressive performance without additional training, several limitations warrant discussion. Firstly, AdaSense's effectiveness is inherently tied to the quality of the posterior sampler, which forms the core of the method. Any limitations in the sampler's capabilities will be reflected in AdaSense's performance. Secondly, generating multiple samples requires high computational cost. Fortunately, the rapid advancements in diffusion-based posterior samplers offer promising avenues for addressing both limitations in the near future. 
Additionally, the current implementation of AdaSense is constrained to linear measurements. 
Exploring the potential of non-linear measurement schemes presents an interesting avenue for future research.

\section{Conclusions}

In this work, we introduced AdaSense, a novel method for adaptive image reconstruction that leverages zero-shot sampling with pre-trained diffusion. AdaSense sequentially refines the reconstruction by strategically selecting the most informative measurements based on the current state. A key strength of AdaSense lies in its versatility. The algorithm remains applicable across diverse image domains, including facial images, MRI, and CT, only adapting the domain-specific diffusion model to capture the unique characteristics of each modality. 

Future work could investigating methods for accelerating AdaSense, perhaps by using previously generated samples. Additionally, it would be interesting to explore extending AdaSense to handle more complex degradation models, such as burst-photography or three-dimension reconstruction with best-next-view. Despite the promising results in the medical setting, we emphasize that more experiments should be conducted before AdaSense can be used in clinical settings.

\par\vfill\par

\section*{Acknowledgements}

This research was partially supported by the Council for Higher Education - Planning \& Budgeting Committee, Israel. This research was also partially supported by the Israel Science Foundation (grant no. 2318/22), by a gift from Elbit Systems, and by the Ollendorff Minerva Center, ECE faculty, Technion.

%
%
\bibliographystyle{splncs04}
\bibliography{refs}
\appendix

\section{Proofs}
\subsection{Proof of \Cref{eq:optimal-constrained}}
\label{app:proof-of-optimal-constrained}
\begin{proof} 
    The solution for $\mV, \mU, \rvb$ in \cref{eq:best-cond-deg} can be viewed as an optimal linear MMSE. The error of this estimator has a closed-form 
    \begin{align}\label{eq:error-linear-estimator} 
        \nonumber 
        \mH_{nr:nr+r}&=\argmin_{\tilde{\mH}} \trace \left\{\mSigma_{\rvx,\rvx|\rvy_{0:nr}} - \mSigma_{\rvx,\tilde{\mH}\rvx|\rvy_{0:nr}}\mSigma_{\tilde{\mH}\rvx,\tilde{\mH}\rvx|\rvy_{0:nr}}^{-1}\mSigma_{\tilde{\mH}\rvx,\rvx|\rvy_{0:nr}}\right\} \\
        \nonumber 
        & \stackrel{1}{=} \argmax_{\tilde{\mH}} \trace \left\{\mSigma_{\rvx,\tilde{\mH}\rvx|\rvy_{0:nr}}\mSigma_{\tilde{\mH}\rvx,\tilde{\mH}\rvx|\rvy_{0:nr}}^{-1}\mSigma_{\tilde{\mH}\rvx,\rvx|\rvy_{0:nr}}\right\} \\
        \nonumber 
        & \stackrel{2}{=} \argmax_{\tilde{\mH}} \trace \left\{\mSigma_{\rvx,\rvx|\rvy_{0:nr}}\tilde{\mH}^\top\left(\tilde{\mH}\mSigma_{\rvx,\rvx|\rvy_{0:nr}}\tilde{\mH}^\top\right)^{-1}\tilde{\mH}\mSigma_{\rvx,\rvx|\rvy_{0:nr}}\right\} \\
        & \stackrel{3}{=} \argmax_{\tilde{\mH}} \trace \left\{\tilde{\mH}\mSigma_{\rvx,\rvx|\rvy_{0:nr}}^{2}\tilde{\mH}^\top\left(\tilde{\mH}\mSigma_{\rvx,\rvx|\rvy_{0:nr}}\tilde{\mH}^\top\right)^{-1}\right\}
    \end{align}
    where $\mSigma_{\rvu,\rvv|\rvy_{0:nr}}$ denotes the covariance matrix $\Cov[\rvu,\rvv|\rvy_{0:nr}]$. \\
    Justifications:
    \begin{enumerate}
        \item $\mSigma_{\rvx,\rvx|\rvy_{0:nr}}$ is not a function of $\tilde{\mH}$, and can be removed from the optimization problem, using the linearity of the trace operator.
        \item $\tilde{\mH}$ is a linear operator, and can apply after the expectation operator used for the covariance.
        \item The trace operator is invariant to cyclic shifts.
    \end{enumerate}
    Constraining $\tilde{\mH}\in\mathcal{H}$ and replacing $\mSigma_{\rvx,\rvx|\rvy_{0:nr}}$ with $\Cov[\rvx|\rvy_{0:nr}]$ in \cref{eq:error-linear-estimator} will complete the proof.
\end{proof}

\subsection{Proof of \Cref{eq:constrained}}
\label{app:proof-of-constrained}
\begin{proof} 
    Starting from \cref{eq:best-cond-deg}, we assume that $\mU=\tilde{\mH}^\dagger$. We first solve the optimization problem for $\tilde{\rvb} = \mV \rvy_{0:nr} +\rvb$,
    \begin{equation} \label{eq:optimal-b}
        \tilde{\rvb} = \argmin_{\tilde{\rvb}} \E \left[ \left\Vert \tilde{\rvb} + \tilde{\mH}^\dagger \tilde{\mH} \rvx - \rvx \right\Vert^2  \middle| \rvy_{0:nr}\right].
    \end{equation}
    The miminum is found when the gradient of \cref{eq:optimal-b} w.r.t. $\tilde{\rvb}$ is zero.
    Therefore, the optimal solution for $\tilde{\rvb}$ is
    \begin{equation} 
        \tilde{\rvb} = (\mI - \tilde{\mH}^\dagger \tilde{\mH})\E \left[\rvx \middle| \rvy_{0:nr}\right].
    \end{equation}
    Next, we are left with finding the optimal $\mH_{nr:nr+r}\in\mathcal{H}$ in the equation
    \begin{align} \label{eq:squared-term}
        \nonumber
        \mH_{nr:nr+r}&=\argmin_{\tilde{\mH}} \E \left[ \left\Vert (\mI - \tilde{\mH}^\dagger \tilde{\mH})\E \left[\rvx \middle| \rvy_{0:nr}\right] + \tilde{\mH}^\dagger \tilde{\mH} \rvx - \rvx \right\Vert^2  \middle| \rvy_{0:nr}\right]\\
        &=\argmin_{\tilde{\mH}} \E \left[ \left\Vert \tilde{\mH}^\dagger \tilde{\mH} (\rvx-\E \left[\rvx \middle| \rvy_{0:nr}\right]) - (\rvx-\E \left[\rvx \middle| \rvy_{0:nr}\right]) \right\Vert^2  \middle| \rvy_{0:nr}\right]
    \end{align}
    Denoting $\bar{\rvx}=\rvx-\E \left[\rvx \middle| \rvy_{0:nr}\right]$, we can rework the squared term in \cref{eq:squared-term},
    \begin{align} \label{eq:open-squared-term}
        \nonumber
        \left\Vert \tilde{\mH}^\dagger \tilde{\mH} \bar{\rvx} - \bar{\rvx} \right\Vert^2
        =\ &\left( \tilde{\mH}^\dagger \tilde{\mH} \bar{\rvx} - \bar{\rvx} \right)^\top \left( \tilde{\mH}^\dagger \tilde{\mH} \bar{\rvx} - \bar{\rvx} \right)\\
        \nonumber
        =\ & \bar{\rvx}^\top \bar{\rvx} - \bar{\rvx}^\top \tilde{\mH}^\dagger \tilde{\mH} \bar{\rvx} - \bar{\rvx}^\top \left(\tilde{\mH}^\dagger \tilde{\mH}\right)^\top \bar{\rvx} + \bar{\rvx}^\top \left(\tilde{\mH}^\dagger \tilde{\mH}\right)^\top \tilde{\mH}^\dagger \tilde{\mH} \bar{\rvx} \\
        \nonumber
        \stackrel{1}{=}\  & \bar{\rvx}^\top \bar{\rvx} - \bar{\rvx}^\top \tilde{\mH}^\dagger \tilde{\mH} \bar{\rvx} - \bar{\rvx}^\top \tilde{\mH}^\dagger \tilde{\mH} \bar{\rvx} + \bar{\rvx}^\top \tilde{\mH}^\dagger \tilde{\mH} \tilde{\mH}^\dagger \tilde{\mH} \bar{\rvx} \\
        \stackrel{2}{=}\  & \bar{\rvx}^\top \bar{\rvx} - \bar{\rvx}^\top \tilde{\mH}^\dagger \tilde{\mH} \bar{\rvx}
    \end{align}
    where (1) and (2) are justified $\left(\tilde{\mH}^\dagger \tilde{\mH}\right)^\top=\tilde{\mH}^\dagger \tilde{\mH}$ and $\tilde{\mH}\tilde{\mH}^\dagger=\mI$ respectively, using the definition of Moore–Penrose pseudo-inverse.
    Combining \cref{eq:squared-term} with \cref{eq:open-squared-term} produces
    \begin{align}
        \nonumber
        \mH_{nr:nr+r}&=\argmin_{\tilde{\mH}} \E \left[ \bar{\rvx}^\top \bar{\rvx} - \bar{\rvx}^\top \tilde{\mH}^\dagger \tilde{\mH} \bar{\rvx}  \middle| \rvy_{0:nr}\right]\\
        \nonumber
        &=\argmax_{\tilde{\mH}} \E \left[\bar{\rvx}^\top \tilde{\mH}^\dagger \tilde{\mH} \bar{\rvx}  \middle| \rvy_{0:nr}\right]\\
        &=\argmax_{\tilde{\mH}} \E \left[(\rvx-\E \left[\rvx \middle| \rvy_{0:nr}\right])^\top \tilde{\mH}^\dagger \tilde{\mH} (\rvx-\E \left[\rvx \middle| \rvy_{0:nr}\right])  \middle| \rvy_{0:nr}\right].
    \end{align}
    Constraining $\tilde{\mH}\in\mathcal{H}$ will complete the proof.
\end{proof}

\section{Considerations for Choosing Posterior Sampler}
\label{app:condiserations-sampler}

While AdaSense, may be used with any posterior sampler, several considerations limit the practical possibilities for posterior sampling methods. Critically, we would prefer the posterior sampler to be a zero-shot method to avoid training for all possible linear degradations. We opt to use a zero-shot diffusion-based inverse problem solver~\cite{kawar2022denoising, song2023pseudoinverse, chung2023diffusion, wang2022zero, mardani2023variational, chung2022improving, graikos2022diffusion}, for which the following properties hold: First, the posterior sampler must be capable of recovering any type of linearly degraded data as long as the degradation $\mH$ is known. Second, $p_\theta(\rvx|\rvy_{0:j})$ must be a stochastic sampler, capable of generating multiple samples instead of a single deterministic output (unless the true distribution $p(\rvx|\rvy_{0:j})$ is a delta function). 

While the results obtained with many such methods are impressive, they often entail high computational demands, requiring between tens to thousands of neural function evaluations. The posterior sampler  used for our method should preferably be computationally efficient, since AdaSense relies on repeated sampling of multiple samples from the posterior distribution. Below, we detail how our choice of DDRM~\cite{kawar2022denoising}, combined with the efficiency considerations offered in \cref{sec:acceleration}, enables us to drastically lower the computational footprint of AdaSense.

DDRM nears the current state-of-the-art performance in inverse problem solving while remaining the on of the fastest method we are aware of~\cite{song2023pseudoinverse, chung2023diffusion, wang2022zero, mardani2023variational, chung2022improving, graikos2022diffusion}, due to the small number of diffusion steps used and the absence of differentiation in the algorithm~\cite{song2023pseudoinverse, chung2023diffusion}. In AdaSense, we limit the number of NFEs to 25 for each posterior sample to balance sampling speed with reconstruction quality.  For the noiseless case and with a specific choice of hyperparameters used in this paper, the DDRM diffusion process can be expressed as  
\begin{equation}
\label{eq:ddrm}
    p_\theta(\rvx_{t-1}|\rvx_t, \rvy) = \normal{( \mI - \mH^\dagger\mH ) \hat{\rvx}_{0}(\rvx_t, t) + \mH^\dagger \rvy}{\sigma_{t-1}^2 \mI}.
\end{equation}
Where $\hat{\rvx}_{0}(\rvx_t, t)$ is the model's inferred prediction for $\rvx_0$ given $\rvx_t, t$, and $\sigma_{t-1}^2$ is the variance of the additive white Gaussian noise at timestep $t-1$. The proof is found in appendix H of~\cite{kawar2022denoising}. This notation highlights DDRM's inherent consistency with the measurement $\rvy$.

\section{The Effect of Sample Size $s$}

\begin{figure}[ht]
  \centering
  \includegraphics[width=0.5\textwidth]{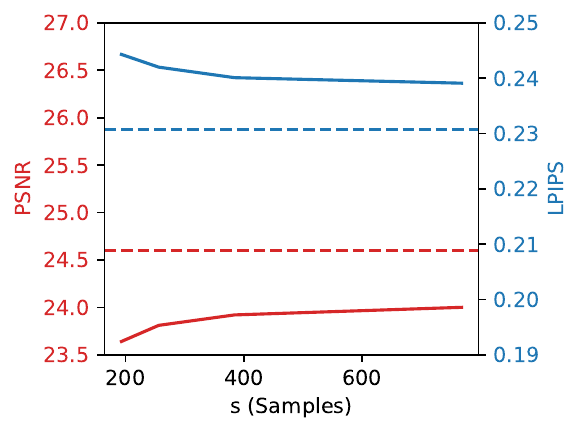}
  \caption{\textbf{Graphs of the effect of samples size $s$ on restoration.} AdaSense is used to restore the same set of images with changing number of posterior samples $s$, while retaining a constant $N=1$, $r=192$. The gain from a very large number of samples is only marginal. The results obtained from using generated samples are similar to ones obtained using real training samples (dotted lines in the right graph). The shown plots are scaled similarly to \cref{fig:adaptivity}.}
  \label{fig:samples}
\end{figure}

To see the effect of the sample size $s$ on the restoration quality, we measure the reconstruction capabilities of AdaSense for different choices of $s$. We set $N=1,r=192$ for simplicity. Using $N=1$ also enables us to compare the results obtained by AdaSense to those achieved by applying PCA on a large number of samples from the training-set, which are not available when considering conditional sampling. In \cref{fig:samples}, we see only marginal increase in performance for much larger sample sizes. Also, results remain close to the upper-bound defined by substituting generated samples for real training data, from which we deduce that using $s\approx\nicefrac{4}{3}\cdot r$ is a good heuristic for an adequate approximation of the real covariance matrix.

\section{Pragmatic Constrained Sampling Considerations}
\label{app:condiserations-constrained}

\subsection{MRI}
In our MRI experiment in \cref{sec:mri} we have used two separate strategies for selecting the subsequent measurement. For the comparison with non-adaptive methods, we apply \cref{eq:constrained} for selecting vertical columns one at a time, enabling parallel exhaustive search over all possible columns. For general MRI subsampling, we select the top $r$ frequencies which maximize \cref{eq:constrained}, therefore also enabling enabling parallel search. In the comparison with adaptive acquisition methods, we mask out all but the central $320\times320$ crop of the pixel-space image, and compute the image intensity before applying \cref{eq:constrained}. This is done to preform a more similar comparison to methods that optimize the squared error of the cropped image intensity. For this reason, as our $\tilde{\mH}\in\mathcal{H}$ apply to complex images, we add an empty imaginary value to the images in the pixel-space.

\begin{figure}[t]
  \centering
  \includegraphics[width=\textwidth]{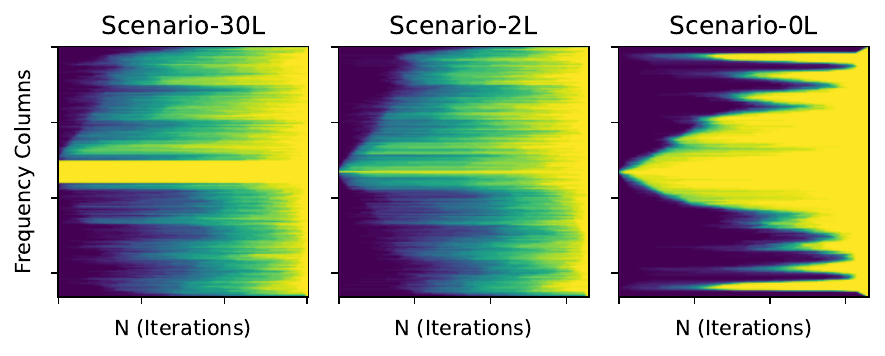}
  \caption{\textbf{MRI acquisition process.} The images depict the probability of choosing a k-space column along the progression of the acquisition process, indicated by the x-axis, when the process runs until all columns have been measured. From left: Scenario-30L and Scenario-2L following \cite{pineda2020active} corresponding to always measuring the lowest 30 or 2 frequencies, respectively, and Scenario-0L corresponding to no pre-set columns as done in \cref{tab:mri-table}.}
    \label{fig:mri-acquisition}
\end{figure}

We visualize the acquisition process for the MRI experiment in \cref{sec:mri} in \cref{fig:mri-acquisition}. Notably while low-frequencies are generally prioritized, several high frequencies also seem to have high importance, especially when not enforcing any low frequencies.

\subsection{CT}
For CT, we pre-compute the matrices corresponding to each row of the Radon transform and use them for applying AdaSense. We also pre-compute $\left(\tilde{\mH}\tilde{\mH}^\top\right)^{-1}$, simplifying the computation of \cref{eq:constrained} during runtime.

\section{Implementation details}
\subsection{AdaSense Hyperparameters}
In \cref{tab:adasense-hyperparameters}, we list our $N,r,s$ hyperparameter choices for all experiments in this work. For unconstrained sampling, we ensure $s > r$ to prevent the selection of degenerate sensing matrices. However, we allow $s > r$ for constrained sampling, as the chosen rows $\mH_{nr:nr+r}$ can rarely span the generated samples $\{\rvx\}_{i=0}^s$ due to the constrains.
\begin{table}[ht]
\centering
\caption{\textbf{AdaSense Hyperparameters}}
    \label{tab:adasense-hyperparameters}
  \centering
  \begin{tabular}{@{}lccc@{}}
    \toprule
    \textbf{Experiment} & \textbf{N} & \textbf{r} & \textbf{s} \\
    \midrule
    \textbf{CelebA} & 8 & 32 & 24 \\
    \textbf{MRI vertical R10} & 33 & 1 & 8 \\
    \textbf{MRI R400} & 16 & 33 & 16 \\
    \textbf{MRI R200} & 32 & 33 & 8 \\
    \textbf{MRI L30 R8} & 12 & 1 & 8 \\
    \textbf{MRI L2 R16} & 19 & 1 & 8 \\
    \textbf{CT $x$ views} & $x$ & 1 & 4 \\
    \bottomrule
  \end{tabular}
\end{table}

\subsection{Diffusion Training}
\begin{table}[ht]
\centering
\caption{\textbf{Diffusion Model Training Hyperparameters}}
    \label{tab:training-hyperparameters}
  \centering
  \begin{tabular}{@{}lcc@{}}
    \toprule
    \textbf{Dataset} & \textbf{FastMRI} & \textbf{DeepLesion} \\
    \midrule
    \textbf{Iterations} & $180$,$000$ & $600$,$000$ \\
    \textbf{Batch Size} & $64$ & $64$ \\
    \textbf{Learning Rate} & $5e-5$ & $2e-4$ \\
    \textbf{Channel Width} & $64$ & $64$ \\
    \textbf{Channel Multipliers} & $[1, 1, 2, 4, 8]$ & $[1, 1, 2, 2, 4, 4]$ \\
    \textbf{Attention Resolutions} & $[23]$ & $[16]$ \\
    \bottomrule
  \end{tabular}
\end{table}
Our experiments were facilitated in part by diffusion models trained for this work, as no pre-trained models where known to us for our data settings for the FastMRI~\cite{knoll2020fastmri, zbontar2018fastmri} or DeepLesion~\cite{yan2018deeplesion} datasets. Our networks use DDPM~\cite{ho2020denoising} U-Net architecture and loss. All networks were trained using the Adam optimizer, dropout with probability $0.1$, and EMA with decay factor of $0.9999$. 
The diffusion process considered in training for all experiments has $1000$ timesteps, with a linear $\beta$ schedule ranging from $\beta_1 = 0.0001$ to $\beta_{1000} = 0.2$. Additional training hyperparameters specific to each experiment can be found in \cref{tab:training-hyperparameters}. All experiments were conducted on $8$ NVIDIA A40 GPUs.


\end{document}